\documentclass[useAMS,usenatbib,fleqn]{mn2e}
\usepackage[dvipdfmx]{graphicx}
 \usepackage{booktabs}
 \usepackage{picins}
  \usepackage{amsmath,amssymb}
  \newcommand{\hM}{\mathcal{M}}
\title[Transonic solutions of isothermal galactic winds]
{Transonic solutions of isothermal galactic winds in a cold dark matter halo}

\author[Tsuchiya, Mori and Nitta]
{Masami Tsuchiya$^{1, 2}$\thanks{
E-mail: mtsuchiya298@gmail.com, mmori@ccs.tsukuba.ac.jp, nittasn@yahoo.co.jp}, Masao Mori$^{2}$ and Shin-ya Nitta$^{3, 4, 5, 6}$\\
$^1$Master's Program in Education, University of Tsukuba, 1-1-1, Tennodai, Tsukuba, Ibaraki, 305-8577, Japan\\
$^2$Center for Computational Sciences, University of Tsukuba, 1-1-1, Tennodai, Tsukuba, Ibaraki, 305-8577, Japan\\
$^3$Research and Support Center on Higher Education for the Hearing and Visually Impaired, Tsukuba University of Technology,\\4-3-15, Amakubo, Tsukuba, Ibaraki, 305-8520, Japan\\
$^4$Hinode Science Project, National Astronomical Observatory of Japan, 
2-21-1 Osawa, Mitaka, Tokyo, 181-8588, Japan\\
$^5$Institute of Space and Astronautical Science, Japan Aerospace Exploration 
Agency, 3-1-1 Yoshinodai, Sagamihara, Kanagawa \\229-8510, Japan\\
$^6$Research Center for Space and Cosmic Evolution, Ehime University, 
2-5 Bunkyo-cho, Matsuyama, Ehime, 790-8577, Japan}

\date{Accepted 2013 April 12.  Received 2013 March 15; in original form 2012 January 20}

\pagerange{\pageref{firstpage}--\pageref{lastpage}} \pubyear{2013}

\begin{document}

\label{firstpage}

\maketitle

\begin{abstract}
We study fundamental properties of steady, spherically symmetric, isothermal galactic outflow in appropriate gravitational potential models. We aim at constructing a universal scale free theory not only for galactic winds, but also for winds from clusters/groups of galaxies. In particular, we consider effects of mass-density distribution on the formation of transonic galactic outflows under several models of the density distribution profile predicted by cosmological simulations of structure formation based on the cold dark matter (CDM) scenario.

In this study, we have clarified that there exists two types of transonic solutions: outflows from the central region and from distant region with a finite radius, depending upon the density distribution of the system. 
The system with sufficiently steep density gradient at the center is allowed to have the transonic outflows from the center. The resultant criterion intriguingly indicates that the density gradient at the center must be steeper than that of the prediction of conventional CDM model including Navarro, Frenk \& White (1997) and Moore et al. (1999). This result suggests that an additional steeper density distribution originated by baryonic systems such as the stellar component and/or the central massive black hole is required to realize transonic outflow from the central region. 
On the other hand, we predict the outflow, which is started at the outskirts of the galactic center and is slowly-accelerated without any drastic energy injection like starburst events. These transonic outflows may contribute secularly to the metal enrichment of the intergalactic medium. 
\end{abstract}


\section{Introduction}

Galactic winds are widely thought to be an essential ingredient in the evolution of galaxies. It plays a key role by which energy and heavy elements are recycled in galaxies and are deposited into the intergalactic space. In the modern paradigm of the galaxy formation based on the cold dark matter (CDM) hypothesis, it deduced that galaxies formed hierarchically in a bottom-up fashion, where a larger system results from the assembly of smaller dark matter halos. Baryonic gas falls into the gravitational potential well of dark matter halos, and condenses rapidly as a result of the radiative cooling for atoms or molecules. The galactic winds seem to be  inevitably influenced by the dark matter gravitational filed. We here study the fundamental nature of the galactic winds in the dark matter gravitational potential.

The existence of heavy elements in the intra-cluster medium of galaxy clusters and the low-density intergalactic medium at $z\sim3$ is the clear evidence of galactic outflows (Songaila 1997; Ellison et al. 2000). Large number of large-scale outflows from star-forming galaxies has been observed in local galaxies (Lehnert \& Heckman 1996; Martin 1999). In the spectra of high-$z$ galaxies, there are evidences for large-scale outflows. For instance, Ly$\alpha$ emission with red asymmetric or P Cygni-type profile is commonly seen in $z>5$ Ly$\alpha$ emitters (Dey et al. 1998; Ellis et al. 2001; Dawson et al. 2002; Ajiki et al. 2002). In addition, extended Ly$\alpha$ nebulae, so called, Ly$\alpha$ blobs at $z>3$ have been possibly interpreted as spatially resolved large-scale winds driven by starbursts (Francis et al. 2001; Ohyama \& Taniguchi 2004; Mori \& Umemura 2006). Lyman break galaxies at $z>3$ also show signs of outflowing gas with velocity of several hundred km s$^{-1}$. Furthermore, Weiner et al. (2009) and Rubin et al. (2010) recently reported that galactic outflows are ubiquitous in the optically selected star-forming galaxies at medium redshift $z \sim 1-2$ using stacked spectra from the DEEP2 survey.

In the theoretical point of view, so far, the large number of studies about the galactic winds has been produced by numerical simulations. The recent developments in computer technologies and numerical methods have made it possible to simulate the multidimensional dynamical or chemodynamical evolution of galaxies including the effect of star formation and supernovae (SNe) feedback. For example, the simulation of the galactic winds from elliptical galaxies are motivated by the observation that elliptical galaxies have very little interstellar medium. While the numerical simulation has become a powerful tool to explore the formation and evolution of galaxies, a simple analytical approach plays an indispensable role to reveal the essential nature of the acceleration process of winds and to interpret observational data. 

Fundamental nature of thermally driven spherically symmetric transonic outflow was examined by Parker (1958) in relation to the solar wind, and then it has been discussed for more than 50 years. This model was based on a hydrodynamic description of the sun's atmosphere, and proposed that the solar wind is a smooth, spherically symmetric, time-steady transonic outflow of hot gas. We must note that the transonic solution is very special and plausible one, because it is the entropy maximum solution connecting the sun and infinity. Thus, such transonic solution is widely accepted as the most plausible universal solution of the outflows from any object. Parker's solution shows that the flow inside the transonic point resembles the hydrostatic equilibrium closely, whereas the flow tends to free-expansion outside the transonic point. 
Then, Holzer and Axford (1970) provided the generalization of Parker's approach to galactic outflow (see also Burke 1968; Chevalier \& Clegg 1985). 
Wang (1995a, 1995b) studied for a radial, steady solution of galactic wind including the effect of the galactic gravitational potential and the efficiency of radiative cooling. 
Assuming a single power-law distribution of mass density, they focused on the outflow that is supersonic everywhere from less massive galaxies.
Recently, Everett \& Murray (2007) investigated the Parker-type acceleration of galactic wind under realistic but complicated situation with gravity from a point mass, cloud/ISM drag force, adiabatic cooling process, and photo ionization/absorption process. They concluded that the outflow accelerated insufficiently under adiabatic cooling.
Moreover, Sharma \& Nath (2012) emphasis that the importance of extra energy source such as photo ionization/absorption process by AGN radiation or SNe energy/momentum injection in order to attain sufficient acceleration.

However, recent cosmological $N$-body simulations based on collisionless CDM scenario no longer indicated such a single power-law distribution but always predicted a double power-law density distribution. 
Navarro, Frenk \& White (1997) pointed out that their structure can be approximated as $\rho_{\rm DM}(r) \propto r^{-1} (r+r_0)^{-2}$, where $r$ is the radius from the center of the galaxy and $r_0$ is the scale radius at which the density profile agrees with the isothermal profile (i.e., $\rho_{\rm DM}(r) \propto r^{-2}$). This is related to the formation epoch of the CDM halo (see also Navarro, Frenk \& White 1997).
Then, Fukushige \& Makino (1997) and Moore et al. (1999) used high-resolution simulations and showed that the CDM halos have a steeper central cusp than quoted above. The resulting structure of the CDM halos depends on the number of particles used in the simulation and is still an open question (see Navarro et al. 2010; Ishiyama et al. 2011). In contrast, recent observations of nearby dwarf galaxies and low surface brightness galaxies have revealed that the density profile of the dark matter halo is constant at the center of such galaxies. For instance, Burkert (1995) proposes that the density profile, $\rho_{\rm DM}(r) \propto (r+r_0)^{-1}(r^2+r_0^2)^{-1}$, nicely reproduces the rotation curves of nearby dwarf galaxies and the central density is correlated with $r_0$ through a simple scaling relation (e.g., Moore 1994; Burkert 1995; de Blok et al. 2001; Swaters et al. 2003; Gentile et al. 2004; Spekkens, Giovanelli \& Haynes 2005). This is well-known as an unsolved problem in the CDM scenario, and so-called the 'Core-Cusp problem'.

Under these inconclusive circumstances for determining CDM distribution, no model considers facing the transonic galactic winds in a realistic constellation of the CDM distribution of galaxies.
These situations motivate us to explore the series of the solution for the transonic galactic winds in the appropriate CDM halo models. In this paper, we focus on the analytical steady-state solution for a spherically symmetric isothermal outflow in a various CDM halo models. We aim at clarifying influence from the dark matter gravitational field on the galactic winds. We are interested especially in realizability of the transonic winds. This research is conducted to be a scale free universal wind theory and does aim not only at winds from galaxies but also at winds from groups/clusters of galaxies because CDM halo density distribution model seems to be rather universal for these hierarchies of objects (Navarro, Frenk \& White 1997). For analytical convenience, we assume isothermal flow without any mass injection along the flow, because gas temperature in actual galaxies seem to be constant (Li et al. 2011) and the locus of the resultant transonic point in this study is far distant from the star forming region (see Fig. \ref{fig2} of this paper). The adequacy of these assumptions is discussed later. 

The structure of this paper is as follows. In Section 2, we describe the basic equations for the analysis of the transonic galactic winds. In Section 3, we show the critical condition for the existence of the transonic winds and their series of solutions. Finally, we discuss the results in Section 4.

\section{Transonic flow in CDM halo}

\subsection{Basic equations}

We consider steady, spherically symmetric, isothermal gas outflows from galaxies. The basic equations for this problem are mass and momentum conservation laws:
\begin{align}
4\pi r^2\rho v =\dot{M}=\mathrm{const},\label{eq:renzoku}
\end{align}
and
\begin{align}
 v\frac{\textrm{d}v}{\textrm{d}r} =-\frac{c_{\rm s}^2}{\rho} \frac{\textrm{d}\rho}{\textrm{d}r}-\frac{\textrm{d}\phi}{\textrm{d}r},\label{eq:undo}
\end{align}
where $\rho , v, {c_\textrm{s}}, \dot{M},$ and $\phi$ are the density, the velocity, the sound speed of the gas, the mass-loss late, and the gravitational potential of a galaxy, respectively. We introduce the nondimensional distance $x=r/r_0$ from the galactic center where $r_0$ is the scale radius of the galaxy. Substituting $\rho$ from equation (\ref{eq:renzoku}) into equation (\ref{eq:undo}), we obtain
\begin{equation}
\frac{c_{\rm s}^2}{\hM}\frac{\textrm{d}\hM}{\textrm{d}x}=\frac{(2c_{\rm s}^2/x)-(\textrm{d}\phi /\textrm{d}x)}{{\hM}^2-1},\label{eq:vkou}
\end{equation}
where $\hM=v/{c_\textrm{s}}$ is the Mach number. The numerator of the right-hand side denotes a change of the effective cross section of a Laval nozzle. Thus, we can see that the gravitational force plays a role of choking of the cross section (hereafter, the gravitational choking). Integrating equation (\ref{eq:vkou}), we obtain the relation,
\begin{equation}
{\hM}^2-\ln{\hM}^2=4\ln x -\frac{2\phi}{c_{\rm s}^2}+C,\label{eq:tokusikiippan}
\end{equation}
where $C$ is the integration constant. A series of wind solutions is obtained from equation (\ref{eq:tokusikiippan}) as ${\hM}^2(x)$ parameterized by $C$.

The feature of solutions of equation (\ref{eq:vkou}) is similar to Parker's solution of solar wind (Parker 1958). The denominator in the right-hand side of equation (\ref{eq:vkou}) is negative for subsonic ($\hM < 1$), and positive for supersonic ($\hM > 1$). If the gravitational choking is effective ($2c_{\rm s}^2/x<\textrm{d}\phi /\textrm{d}x)$, the numerator is negative, while if the gravitational choking is not effective ($2c_{\rm s}^2/x>\textrm{d}\phi /\textrm{d}x)$, the numerator is positive. It is equivalent to the cross section of a Laval nozzle that is decreasing (increasing) for negative (positive) numerator. Subsonic flow accelerates in the region where the numerator is negative and supersonic flow accelerates in the region where the numerator is positive. The singular point at which both the denominator and the numerator equal to zero ($\hM=1,2c_{\rm s}^2/x=\textrm{d}\phi /\textrm{d}x$), and sign of the numerator changes from negative to positive is called the transonic point(or the critical point, see Bondi 1952; Parker 1958). Only the transonic flow can continuously accelerate from subsonic to supersonic by passing through the transonic point.

\subsection{Density distributions}

The NFW model that is widely accepted density profile of CDM halos was empirically derived from cosmological $N$-body simulation (Navarro, Frenk \& White 1997). Fukushige \& Makino (1997) and Moore et al. (1999) derived steeper distribution from $N$-body simulation with higher resolution.
The resultant empirical profiles of density distribution is approximately fitted by
\begin{equation}
\rho_\textrm{\tiny DM}(r;\alpha )= \frac{\rho_0{r_0}^3}{r^\alpha (r+r_0)^{3-\alpha}},\label{eq:densityr}
\end{equation}
where $\rho_0$ is the scale density. In this equation, $\alpha=1$ is the NFW model and $\alpha=1.5$ corresponds to the model advocated in Fukushige \& Makino (1997) and Moore et al. (1999). On the other hand, observed density profile derived by Moore (1994) and Burkert (1995) is represented by $\alpha=0$. The plausible value of the index $\alpha$ remains an open question. Thus, we treat $\alpha$ as a parameter and study the variation of solutions depending on $\alpha$. By choosing $r_0$ as the unit of length, equation (\ref{eq:densityr}) is rewritten as
\begin{equation}
\rho_\textrm{\tiny DM}(x;\alpha )= \frac{\rho_0}{x^\alpha (x+1)^{3-\alpha}}, \label{eq:density}
\end{equation}
where $x=r/r_0$. In the limit of $x\rightarrow 0$, $\rho_\textrm{\tiny DM}(x;\alpha ) \propto x^{-\alpha}$ and $\rho_\textrm{\tiny DM}(x;\alpha ) \propto x^{-3}$ for $x\rightarrow \infty $. Note that the functional form of the gravitational potential near the center crucially depends on the index $\alpha$.

\section{Series of galactic wind solutions}

\subsection{Condition for transonic flow from the center}

The acceleration of outflows obviously depends on the shape of the gravitational potential from the discussion in section 2.1, and the existence of transonic solutions starting from the center, therefore, relates closely to the power-law index $\alpha$ in equation (\ref{eq:densityr}). Here, we consider the critical condition upon $\alpha$ for the transonic solution from the center.

The total mass $M(x)$ within $x$ under the density distribution (\ref{eq:density}) can be described analytically using Gauss's hypergeometric function (see appendix A) as
\begin{equation}
M(x;\alpha )= \frac{4\pi \rho_0{r_0}^3}{3-\alpha} x^{3-\alpha} \,_2\mathrm{F}_1 \left[ 3-\alpha,3-\alpha,4-\alpha ;-x \right],
\label{eq:mass}
\end{equation}
if $0\leq \alpha<3$. Substituting $\textrm{d}\phi (x;\alpha )/\textrm{d}x=GM(x;\alpha )/(r_0x^2)$ into the numerator of the right-hand side of equation (\ref{eq:vkou}), we obtain
\begin{equation}
\frac{2c_{\rm s}^2}{x}-\frac{\textrm{d}\phi (x;\alpha )}{\textrm{d}x}=\frac{2c_{\rm s}^2}{x} \left[ 1-\sum^{\infty}_{n=0}A_n(-1)^nx^{n+2-\alpha} \right],\label{eq:senjouken}
\end{equation}
where \{$A_n$\} are positive constants (see appendix A). 
In the limit of $x\rightarrow 0$, equation (\ref{eq:senjouken}) approaches
\begin{equation}
\frac{2c_{\rm s}^2}{x}-\frac{\textrm{d}\phi (x;\alpha )}{\textrm{d}x}\simeq \frac{2c_{\rm s}^2}{x} \left( 1-A_0x^{2-\alpha} \right).\label{eq:senjoukenkunji}
\end{equation}
If the choking by the gravitational force is effective in the innermost region, the numerator is negative. In the case of $0\leq \alpha<2$, the numerator is positive because $ 1-A_0x^{2-\alpha} \rightarrow 1$ as $x \rightarrow 0$.
In the case of $\alpha=2$, $ 1-A_0x^{2-\alpha} = 1-A_0$. Therefore the numerator is negative if $A_0>1$. In the case of $2< \alpha <3$, the numerator is negative because $1-A_0x^{2-\alpha} \rightarrow -\infty $ as $x \rightarrow 0$. Finally, we conclude that the condition for realizing transonic flow from the central region is $\alpha\geq 2$.

We note that this is rather steep density profile in the innermost region. Widely accepted values of $\alpha$ for the density distribution profile of CDM halos are $\alpha=$ 1 and 1.5, but they do not fulfill the critical condition $\alpha\geq 2$. The galactic wind solutions for the case of $0\leq \alpha<2$ (Case 1), and the case of $2\leq \alpha \leq 3$ (Case 2) are shown in the following sections. Here, we focus only on the cases $\alpha=0, 1$ and 1.5 in Case 1, and $\alpha=$ 2 and 2.5 in Case 2 because of analytical simplicity.

\subsection{Case 1: $0\leq \alpha<2$}

\subsubsection{The galaxy mass, the gravitational force, and the gravitational potential}
The total mass $M(x;\alpha$) within $x$ is given for specified values of $\alpha$ as follows,
\begin{align}
M(x;\alpha =0)&=4\pi \rho_0{r_0}^3 \left[ \ln(x+1)-\frac{x(2+3x)}{2(x+1)^2} \right],\\
M(x;\alpha =1)&= 4\pi \rho_0{r_0}^3 \left[ \ln(x+1)-\frac{x}{x+1}\right], \label{eq:situnfw}
\end{align}
and
\begin{align}
M(x;\alpha =1.5)&=8\pi \rho_0{r_0}^3 \nonumber \\
&\left[ \ln(\sqrt{x}+\sqrt{x+1})-\sqrt{\frac{x}{x+1}} \right].
\end{align}
Therefore the corresponding gravitational forces are given by
\begin{align}
\frac{\textrm{d}\phi (x;\alpha =0)}{\textrm{d}x}&=4\pi \rho_0 {r_0}^2 G \nonumber \\
&\frac{1}{x^2} \left[ \ln(x+1)-\frac{x(2+3x)}{2(x+1)^2} \right], \label{eq:jyuryokua0}\\
\frac{\textrm{d}\phi (x;\alpha =1)}{\textrm{d}x}&=4\pi \rho_0{r_0}^2G\frac{1}{x^2} \left[ \ln(x+1)-\frac{x}{x+1}\right], \label{eq:jyuryokuNFW}
\end{align}
and
\begin{align}
\frac{\textrm{d}\phi (x;\alpha =1.5)}{\textrm{d}x}&=8\pi \rho_0 {r_0}^2 G \nonumber \\
&\frac{1}{x^2} \left[ \ln(\sqrt{x}+\sqrt{x+1})-\sqrt{\frac{x}{x+1}} \right], \label{eq:jyuryokua1.5}
\end{align}
respectively.
By integrating equations (\ref{eq:jyuryokua0}), (\ref{eq:jyuryokuNFW}) and (\ref{eq:jyuryokua1.5}), corresponding gravitational potentials are given as
\begin{align}
\phi (x;\alpha =0)&= -4\pi\rho_0 {r_0}^2G \nonumber \\
&\left[ \frac{1}{x}\ln(x+1)-\frac{1}{2(x+1)} \right], \label{eq:poten0}\\
\phi (x;\alpha =1) &= -4\pi\rho_0 {r_0}^2G\frac{1}{x}\ln(x+1), \label{eq:poten1}
\end{align}
and
\begin{align}
\phi (x;\alpha =1.5) &= -8\pi\rho_0 {r_0}^2G \nonumber \\
&\frac{1}{x} \left[ 1+\ln(\sqrt{x}+\sqrt{x+1}-\sqrt{x}\sqrt{x+1}) \right], \label{eq:poten1.5}
\end{align}
respectively.

\subsubsection{Mechanism of acceleration}
\begin{figure}
\begin{center}
\includegraphics[width=8cm,clip]{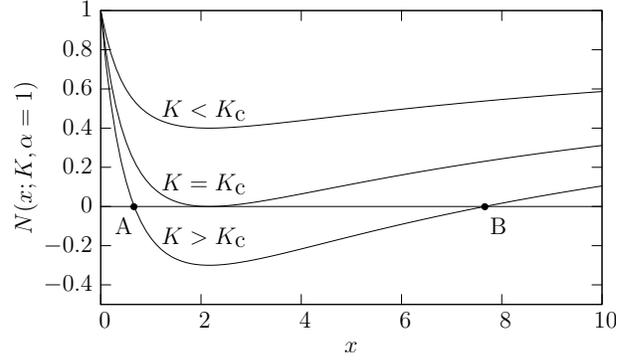}
\caption{$N(x;K, \alpha)$ with $\alpha=1$: If $K>K_\textrm{c}$, two points (A, B) at which $N(x;K, \alpha)$ vanishes appear. The gravitational choking is effective in $x_\textrm{A}<x<x_\textrm{B}$ ($x_\textrm{A},x_\textrm{B}$ are the loci of A and B, respectively). This pattern does not topologically change in $0\leq \alpha<2$.}
\label{fig1}
\end{center}
\end{figure}

From the numerator of right-hand side of equation (\ref{eq:vkou}), we define
\begin{equation}
\frac{2c_{\rm s}^2}{x}-\frac{\textrm{d}\phi (x;\alpha)}{\textrm{d}x}\equiv \frac{2c_{\rm s}^2}{x}N(x;K, \alpha), \label{eq:gravsexpan}
\end{equation}
\begin{align}
N(x;K, \alpha)&=1-Kx^{2-\alpha}\frac{\,_2\mathrm{F}_1 \left[ 3-\alpha,3-\alpha;4-\alpha;-x \right]}{3-\alpha} \label{numerator},
\end{align}
and
\begin{align}
K&=\frac{2\pi\rho_0 {r_0}^2G}{c_{\rm s}^2}.\label{eq:K}
\end{align}
Substituting equations (\ref{eq:jyuryokua0}), (\ref{eq:jyuryokuNFW}) or (\ref{eq:jyuryokua1.5}), in sequence, into the numerator of the right-hand side of equation (\ref{eq:vkou}), we obtain
\begin{align}
N(x;K, \alpha=0)&=1-\frac{K}{x} \left[ \ln(x+1)-\frac{x(2+3x)}{2(x+1)^2} \right], \label{eq:fa=0}\\
N(x;K, \alpha=1)&=1-\frac{K}{x}\left[\ln(x+1) -\frac{x}{x+1}\right], \label{eq:fa=1}
\end{align}
and
\begin{align}
N(x;K, \alpha=1.5)&=1-\frac{2K}{x} \nonumber \\
&\left[ \ln(\sqrt{x}+\sqrt{x+1})-\sqrt{\frac{x}{x+1}} \right], \label{eq:fa=1.5}
\end{align}
for each value of $\alpha$. If the factor $N(x;K, \alpha)$ is positive, subsonic flow decelerates and supersonic flow accelerates, while if $N(x;K, \alpha)$ is negative, subsonic flow accelerates and supersonic flow decelerates. Fig. \ref{fig1} shows the behavior of the factor $N(x;K, \alpha)$ with $\alpha=1$ for example. We can find that there are three cases depending on $K$.  
\begin{figure}
\begin{center}
\includegraphics[width=8cm]{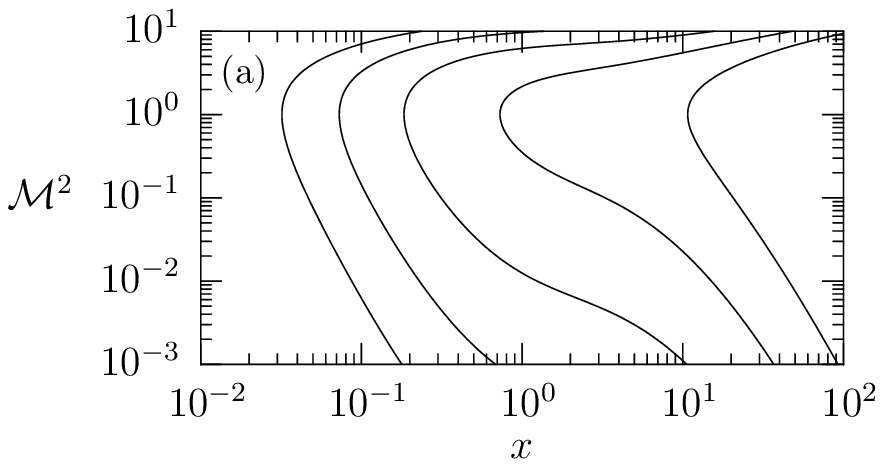}\\
\includegraphics[width=8cm]{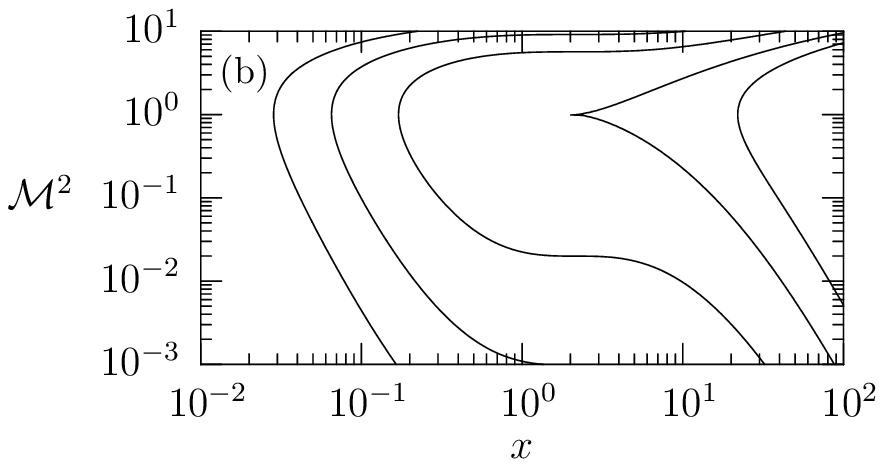}\\
\includegraphics[width=8cm]{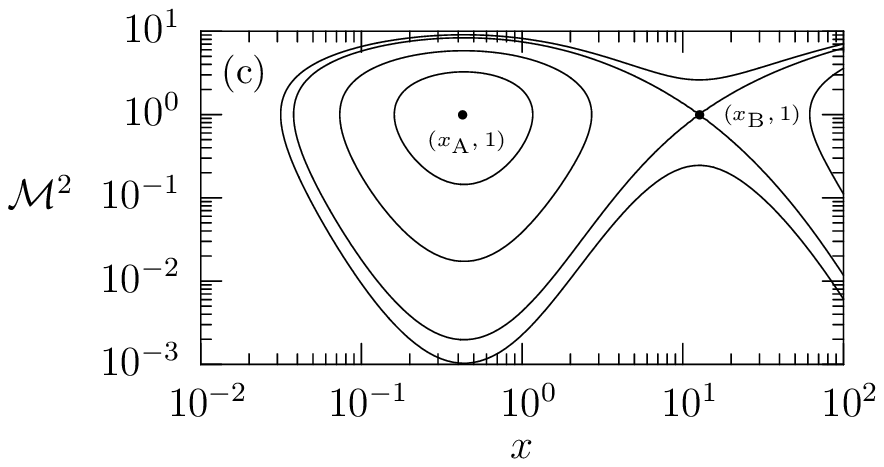}
\caption{Family of solutions for $\alpha=1$: The upper panel is for $K=3.8(<{K_\textrm{c}}$), the middle panel and the lower panel are for $K={K_\textrm{c}} \simeq 4.6$, and for $K=7.5(>{K_\textrm{c}})$, respectively. In lower panel, sign of velocity gradient changes at $x_\textrm{A} \simeq 0.43$ and $x_\textrm{B} \simeq 12.7$ and transonic solution which passes the transonic point $(x,{\hM}^2)\simeq (12.7, 1)$ appears. The patterns of solution curves for $0\leq \alpha<2$ do not topologically change.}
\label{fig2}
\end{center}
\end{figure}
A and B in Fig. \ref{fig1} are the points at which $N(x;K, \alpha)$ vanishes.
There is a critical value $K={K_\textrm{c}}$ such that minimum of $N(x;K_\textrm{c}, \alpha)$ is zero. The value of ${K_\textrm{c}}$ depends on $\alpha$ as follows: ${K_\textrm{c}} \simeq 4.6$ for $\alpha =1$, ${K_\textrm{c}} \simeq 2.9$ for $\alpha =1.5$, and ${K_\textrm{c}}\simeq 8.2$ for $\alpha =0$. If $K<{K_\textrm{c}}$, $N(x;K, \alpha)$ is always positive. If $K>{K_\textrm{c}}$, two points (A, B) at which $N(x;K, \alpha)$ vanishes appear. We name the loci of A and B as $x_\textrm{A}$ and $x_\textrm{B}$, respectively. $x_\textrm{A}$ is the locus that sign of the factor $N(x;K, \alpha)$ changes from positive to negative. $x_\textrm{B}$ is the locus that sign of the factor $N(x;K, \alpha)$ changes from negative to positive. Thus, the point $(x,{\hM}^2)=( x_\textrm{B},1)$ at which galactic wind passes $x_\textrm{B}$ with sound speed become the transonic point.

\subsubsection{Topology of the solution}
We substitute equations (\ref{eq:poten0}), (\ref{eq:poten1}), and (\ref{eq:poten1.5}) into equation (\ref{eq:tokusikiippan}) successively and draw solutions in the plane ($x,{\hM}^2$) for each case of $\alpha$. Fig. \ref{fig2} shows family of solutions for $\alpha=1$ for example. Fig. \ref{fig2} is the so-called `phase diagram' of the solutions of equation (\ref{eq:tokusikiippan}). We can see that the topological property of the phase diagram strongly depends on $K$ as shown in Fig. \ref{fig1}. Fig. \ref{fig2}(a), \ref{fig2}(b) and \ref{fig2}(c) are for $K=3.8(<{K_\textrm{c}})$, $K={K_\textrm{c}}\simeq 4.6$ and $K=7.5(>{K_\textrm{c}})$, respectively.

If $K\leq {K_\textrm{c}}$, subsonic flow decelerates and supersonic accelerates monotonically (Fig. \ref{fig2}(a) and \ref{fig2}b). Consequently, transonic solution does not appear in these cases. If $K>{K_\textrm{c}}$, the subsonic flow accelerates and supersonic flow decelerates in $x_\textrm{A}<x<x_\textrm{B}$ (Fig. \ref{fig2}c). The point at $(x,{\hM}^2)=( x_\textrm{A},1)$ is so-called `O-point' from its topological property in Fig. \ref{fig2}(c). Similarly, the point at $(x,{\hM}^2)=( x_\textrm{B},1)$ is so-called `X-point'. This is the transonic point. In Fig. \ref{fig2}(c), the sign of velocity gradient changes at $x_\textrm{A} \simeq 0.43$ and $x_\textrm{B} \simeq 12.7$. The transonic point forms at $(x,{\hM}^2)\simeq (12.7, 1)$. Therefore a transonic solution which passes $(x,{\hM}^2)\simeq (12.7, 1)$ exists.

\subsection{Case 2: $2\leq \alpha <3$}

\subsubsection{The galaxy mass, the gravitational force, and the gravitational potential}

The total mass $M(x;\alpha$) within $x$ for $\alpha=2$ and $2.5$ are given as
\begin{align}
M(x;\alpha =2)&= 4\pi \rho_0{r_0}^3 \ln(x+1), 
\end{align}
and
\begin{align}
M(x;\alpha =2.5)&=8\pi \rho_0{r_0}^3 \ln\left(\sqrt{x}+\sqrt{x+1}\right),
\end{align}
respectively. The gravitational forces for each case, therefore, are
\begin{align}
\frac{\textrm{d}\phi (x;\alpha =2)}{\textrm{d}x}&=4\pi \rho_0{r_0}^2G\frac{1}{x^2} \ln(x+1), \label{eq:jyuryokua2} 
\end{align}
and
\begin{align}
\frac{\textrm{d}\phi (x;\alpha =2.5)}{\textrm{d}x}&=8\pi \rho_0 {r_0}^2 G\frac{1}{x^2}\ln\left(\sqrt{x}+\sqrt{x+1}\right), \label{eq:jyuryokua2.5}
\end{align}
respectively.
By integrating equation (\ref{eq:jyuryokua2}) and (\ref{eq:jyuryokua2.5}), the gravitational potentials for each case are given as
\begin{align}
\phi (x;\alpha &=2) = -4\pi\rho_0 {r_0}^2G \nonumber \\
&\left[ \frac{1}{x}\ln(x+1)-\ln \frac{x}{x+1}\right], \label{eq:poten2} 
\end{align}
and
\begin{align}
\phi (x;\alpha &=2.5) = -8\pi\rho_0 {r_0}^2G \nonumber \\
&\left[\frac{\sqrt{x}\sqrt{x+1}+\ln\left(\sqrt{x}+\sqrt{x+1}\right)}{x}-1\right],  \label{eq:poten2.5},
\end{align}
respectively.

\subsubsection{Mechanism of acceleration}
\begin{figure}
\begin{center}
\includegraphics[width=8cm,clip]{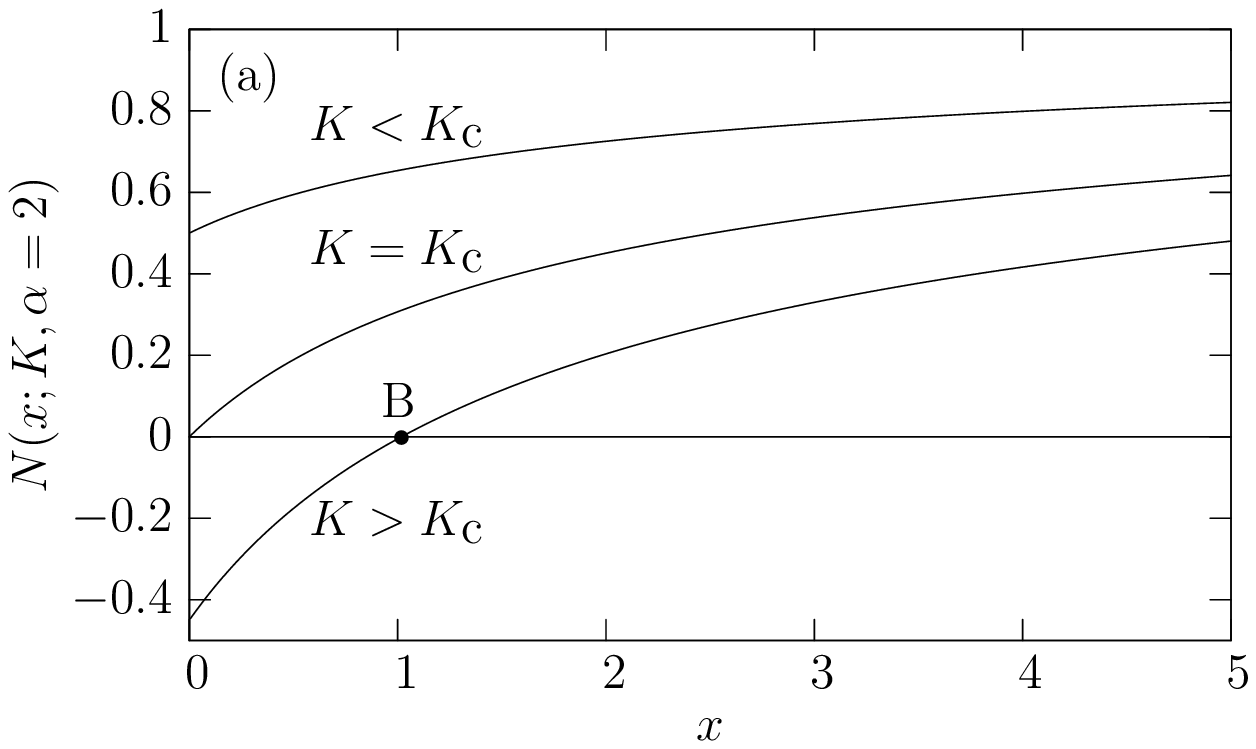}\\
\includegraphics[width=8cm,clip]{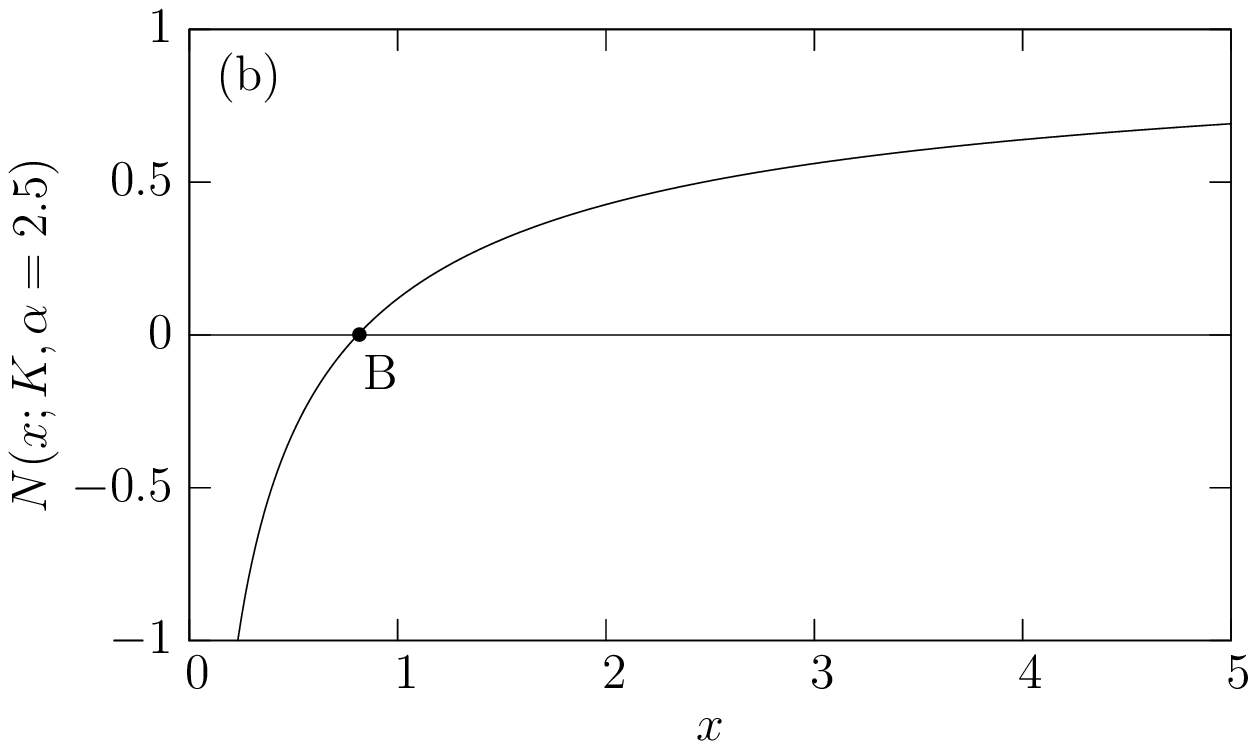}
\caption{$N(x;K, \alpha)$ with $\alpha=2$ (upper panel) and with $\alpha=2.5$ (lower panel), respectively. Lower panel is for $K=0.50$. The pattern of lower panel does not topologically change in $2\leq \alpha <3$.}
\label{fig3}
\end{center}
\end{figure}
Substituting equations (\ref{eq:jyuryokua2}) or (\ref{eq:jyuryokua2.5}) into the numerator of the right-hand side of (\ref{eq:vkou}), we obtain
\begin{align}
N(x;K, \alpha=2)&=\frac{K}{x}\ln(x+1), \label{eq:fa=2} 
\end{align}
and
\begin{align}
N(x;K, \alpha=2.5)&=\frac{2K}{x}\ln\left(\sqrt{x}+\sqrt{x+1}\right). \label{eq:fa=2.5} 
\end{align}
Fig. \ref{fig3}(a) shows $N(x;K, \alpha)$ for $\alpha=2$ and Fig. \ref{fig3}(b) shows $\alpha=2.5$ as a function of $x$. We can see that the function $N(x;K, \alpha)$ monotonically increases for $\alpha \geq 2$: it is critically different from Case 1. In the Case 2, one X-point without O-point appears. In this case, we define B as shown in Fig. \ref{fig3}. In the case of $\alpha=2$, the value of $N(x;K, \alpha)$ is finite at $x=0$ and there is a critical value $K=K_\textrm{c}=1$. Therefore if $K\leq 1$, $N(x;K, \alpha)$ does not become negative, while if $K>1$, $N(x;K, \alpha)$ becomes negative in $x<x_\textrm{B}$ ($x_\textrm{B}$ is the locus of B) in Fig. \ref{fig3}(a). Thus, the point where $(x,{\hM}^2)=( x_\textrm{B},1)$ becomes the transonic point. In the case of $\alpha=2.5$, the value of the function is negative infinity in the limit of $x\rightarrow 0$. Therefore a transonic point always exists at $(x,{\hM}^2)=( x_\textrm{B},1)$ in Fig. \ref{fig3}(b).

\subsubsection{Topology of the solution}

\begin{figure}
\begin{center}
\includegraphics[width=8cm]{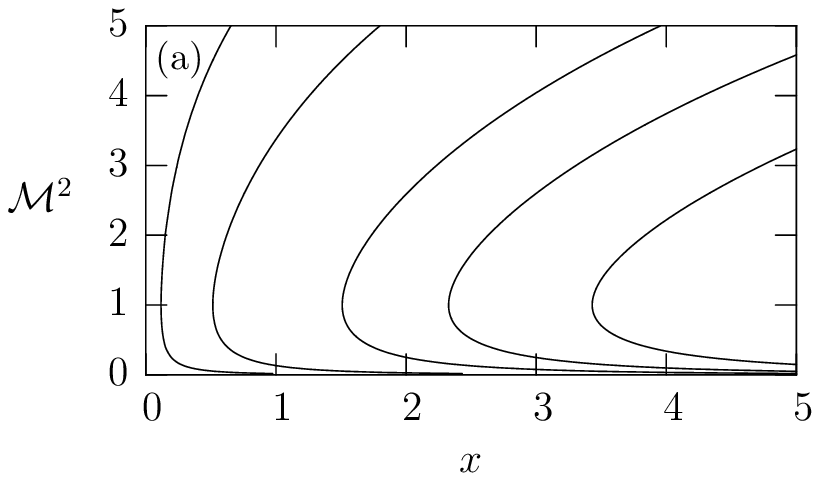}\\
\includegraphics[width=8cm]{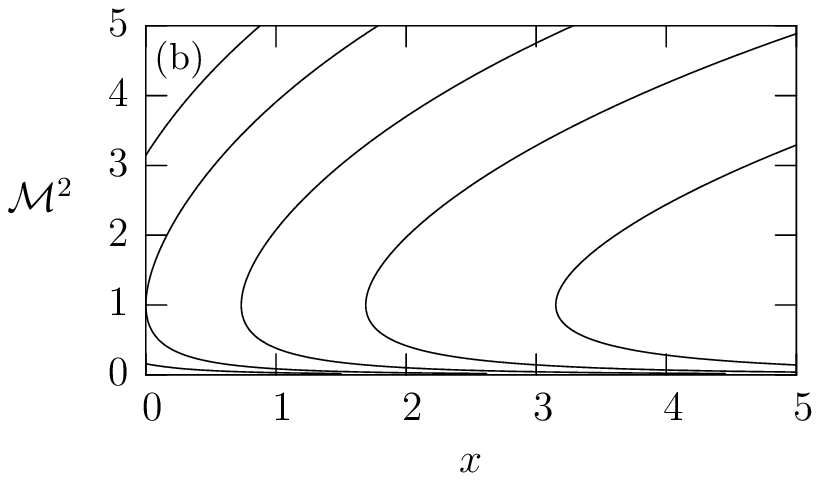}\\
\includegraphics[width=8cm]{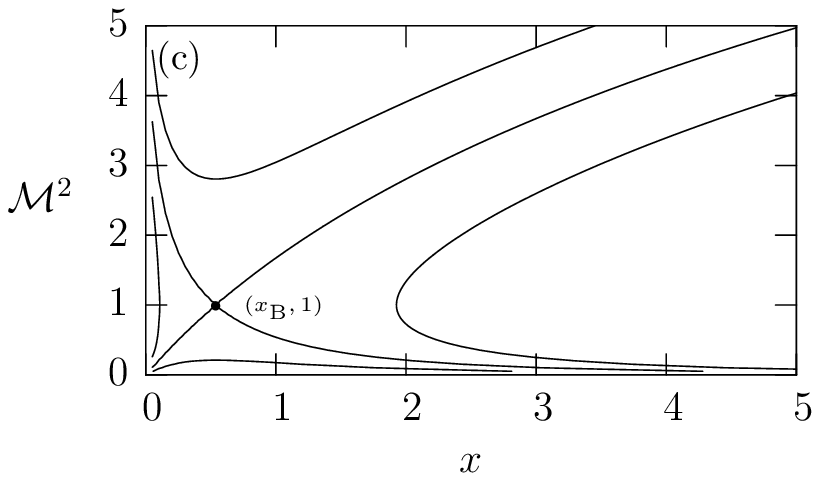}
\caption{Family of solutions for $\alpha=2$: The upper panel is for $K=0.75<{K_\textrm{c}}$, the middle panel is for $K={K_\textrm{c}}=1$, and the lower panel is for $K=1.25>{K_\textrm{c}}$. In lower panel, sign of velocity gradient changes at $x_\textrm{B}\simeq 0.54$ and transonic solution which passes the transonic point $(x,{\hM}^2)\simeq (0.54, 1)$ appears.}
\label{fig4}
\end{center}
\end{figure}

\begin{figure}
\begin{center}
\includegraphics[width=8cm]{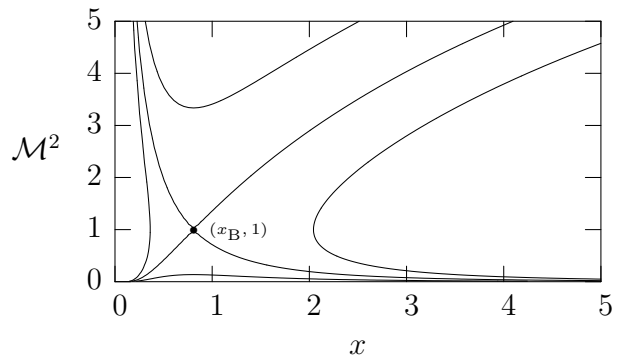}
\caption{Family of solutions for $\alpha=2$, for $K=0.50$: Sign of velocity gradient changes at $x_\textrm{B}\simeq 0.81$ and transonic solution which passes the transonic point $(x,{\hM}^2)\simeq (0.81, 1)$ appears. The patterns of solution curves for $0\leq \alpha<2$ do not topologically change.}
\label{fig5}
\end{center}
\end{figure}

By substituting equations (\ref{eq:poten2}) or (\ref{eq:poten2.5}) into equation (\ref{eq:tokusikiippan}), we can draw phase diagram for each case. Figs. \ref{fig4} and \ref{fig5} are phase diagram for $\alpha=2$ and $\alpha=2.5$, respectively. 
Figs. \ref{fig4}(a), (b) and (c) are for $K=0.75(<{K_\textrm{c}})$, $K=1(={K_\textrm{c}})$, $K=1.25(>{K_\textrm{c}})$, and Fig. \ref{fig5} is for $K=0.50$, respectively. In the case of $\alpha=2$ (Fig. \ref{fig4}), if $K>{K_\textrm{c}}$, $x_\textrm{B}$ at which the sign of the velocity gradient changes appears and there is a transonic solution which passes the transonic point at $(x,{\hM}^2)=(x_\textrm{B},1)$. In the case of $\alpha=2.5$ (Fig. \ref{fig5}), there is also a transonic solution which passes the transonic point at $(x,{\hM}^2)=(x_\textrm{B},1)$. In this case, according to Section 3.3.2, we find such critical solution for any value of $K$. According to Figs. \ref{fig4}(c) and \ref{fig5}, the transonic wind can start from the center $x=0$ in Case 2.

\section{Summary and Discussion}

The property of a stationary, spherically symmetric and isothermal galactic wind in a CDM halo is examined in this paper. Depending on the mass-density profile of the CDM halo, three types of outflow can exist: (i) supersonic or subsonic flows everywhere, and flows with (ii) an X-point or (iii) a pair of an O-point and an X-point. The condition for existence of a transonic solution is sensitive to the mass-density profile of the CDM halo. Especially, the transonic outflow from the galactic center is realized under necessary condition that the power-law index of the mass-density distibution near the galactic center must be steeper than two ($\alpha\geq 2$ in equation \ref{eq:densityr}).

\subsection{The locus of the transonic point}

\begin{figure}
\begin{center}
\includegraphics[width=8cm,clip]{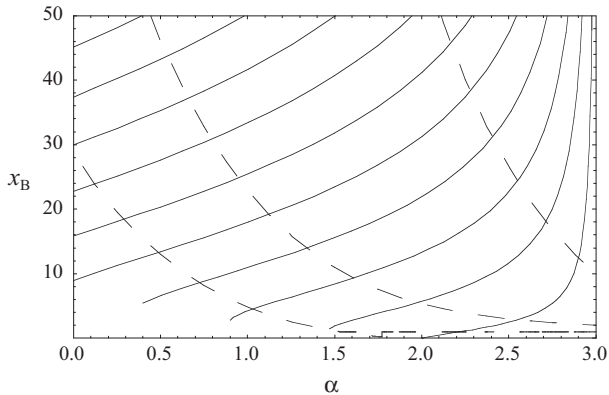}
\caption{Locus of the outer X-point $x_{\rm c}$ obtained by solving equation $N(x_{\rm c}; K, \alpha)=0$ for $0 \leq \alpha < 3$ and $ 1 \leq K \leq 19$. Each solid-curve corresponds $K=1, 3, 5, 7, 9, 11, 13, 15, 17$ and 19 from bottom to top. Each dashed-curve corresponds $K=f(\alpha)/\gamma$ with $\gamma=1, 0.5$ and 0.1 from bottom to top.}
\label{fig6}
\end{center}
\end{figure}

Our analysis shows that the locus of the transonic point is critically related to both the power-law index $\alpha$ of the mass-density distribution and the coefficient $K$ (see equations \ref{eq:density} and \ref{eq:K}). In Fig. \ref{fig6}, we summarize the locus $x_{\rm B}$ of the transonic point obtained by solving equation $N(x_{\rm B}; K, \alpha)=0$ for $0 \leq \alpha < 3$ and $ 1 \leq K \leq 19$. Each solid-curve shows the locus $x_{\rm B}$ as function of $\alpha$ for given value of $K$ in the range from $K=1$ (bottom) to 19 (top). Using equation (\ref{eq:mass}) and $c_{\rm s}^2 = k_{\rm B} T / \mu m_{\rm p}$, where $\mu$ is the mean molecular weight, $m_{\rm p}$ is the proton mass, $k_{\rm B}$ is Boltzmann's constant, and $T$ is the gas temperature, equation (\ref{eq:K}) becomes
\begin{equation}
K = \frac{G M(r_0; \alpha) \mu m_{\rm p}}{2 r_0  k_{\rm B} T} f(\alpha), \label{eq:KusingT}
\end{equation}
where $M(r_0; \alpha)$ is the enclosed mass within the radius $r=r_0$ for a given $\alpha$, and $f(\alpha)$ is defined by
\begin{equation}
f(\alpha) = \frac{3-\alpha}{_2\mathrm{F}_1 \left[ 3-\alpha,3-\alpha,4-\alpha ;-1 \right]}.
\end{equation}
The numerical values of $f(\alpha)$ for typical $\alpha$ are summarized in Table \ref{table:f}. Accordingly, the coefficient $K$ is roughly interpreted as the ratio of the gravitational potential energy of the galaxy to the thermal energy of the gas.

%----------- fig. 6 -----------% 
%------------------------------% 

%---------- Table 1 -----------% 
\begin{table}
\caption{Numerical values of $f(\alpha)$ for typical $\alpha$}
\label{table:f}
\begin{tabular}{@{}lccccccc}
\hline
$\alpha$ & 0 & 0.5 & 1.0 & 1.5 & 2.0 & 2.5 & 3.0 \\
\hline
$f(\alpha)$ & 14.7 & 8.86 & 5.18 & 2.87 & 1.44 & 0.567 & 0 \\
\hline
\end{tabular}
\end{table}
%------------------------------% 

Fig. \ref{fig6} indicates the allowable range of the transonic point for given $K$. For instance, the transonic solution can exist only in the range $2 \leq \alpha < 3 $ for $K=1$ and $0 \leq \alpha <3$ for $K > 8.15$. Especially, the transonic solution starting from the center $x=0$ can exist only for $\alpha \geq 2$. This is consistent with the discussion in section 3.1. In addition, we can see that the locus of the transonic point is an increasing function of $K$ for given $\alpha$. For the NFW model ($\alpha=1$), the transonic point locates at $x_{\rm B}=4.22$ for $K=5$ and $x_{\rm B}=21.6$ for $K=10$. Incidentally, the virial radius of the CDM halo is usually defined by $r_{\rm vir} = c \, r_{\rm 0}$ for the NFW model, where $c$ is so-called the concentration parameter. Using the concentration-mass relation 
\begin{equation}
\log_{10}c=0.971-0.094\log_{10}(M/10^{12}h^{-1}\textrm{M}_\odot ),
\end{equation}
derived by Maccio, Dutton \& van den Bosch (2008), we obtain $c$ as a function of the virial mass. This equation gives $ 7 \la c \la 30 $ with respect to a reasonable range of the virial mass $7 \la \log_{10}(M/\textrm{M}_\odot ) \la 13$ for the NFW model. In other words, the virial radius is in the range $ 7 \, r_0 \la r_{\rm vir} \la 30 \, r_0$. Note that the virial radius is comparable to the locus of the transonic point for $\alpha=1$ in Fig. 6. Depending on the coefficient $K$ and the virial mass of the galaxies, therefore, the wind is accelerated mainly in the region far outside the optically visible scale of the galaxies in some cases. Moreover, we must note that the density structure of the transonic wind in the subsonic region $x < x_{\rm B}$ is crucially similar to that of the hydrostatic equilibrium except in the vicinity of the transonic point. Thus, we may observationally confound the slowly accelerated wind structure having $x_{\rm B} r_0 \gg r_{\rm vir}$ with the hydrostatic one. 

There possibly exists such slowly accelerated outflows in the outskirts of galaxies with no drastic heating by such as starburst events but by quasi-stationary heating from the heat reservoir of shock heated CDM halo (see next subsection). Such slowly accelerated galactic wind may play an important role of the metal enrichment of the intergalactic medium though it resembles closely to the hydrostatic gas in the optically observable region.

\subsection{Availability of our simplified model}

\subsubsection{Isothermal approximation}
\label{sec: isithermal-approx}

In the standard CDM picture of galaxy formation, gas falling into CDM halos is shock-heated approximately to the halo virial temperature, thus the CDM halo maintains quasi-hydrostatic equilibrium (Rees \& Ostriker 1977; Binney 1977; Silk 1977; White \& Rees 1978). We assume that the wind gas is in thermally equilibrium with such heated CDM halo. In this model, CDM halo plays a role of a heat reservoir to keep wind gas temperature nearly to be the halo virial temperature. In this case, the gas could be treated as isothermal and its temperature $T$ is expected to be close to the virial temperature in a wide range up to the virial radius, 
\begin{equation}
T_{\rm vir} =  \gamma \frac{G \mu m_{\rm p} M(r_0; \alpha)}{2 k_{\rm B} r_0},
\end{equation}
where $\gamma$ is a fudge factor of the order of unity that should be determined by the efficiency of the shock heating. A typical value of $T_{\rm vir}$ is $\sim$ several$\times 10^6$ K for galaxies with mass $\sim 10^{11}$ M$_\odot$ and scale radius $\sim 10$ kpc. 

Indeed, in case of the Sombrero galaxy, the gas temperature seems to be isothermal in a considerably wide spatial range $\leq$ 25 kpc from the galactic center (Li et al. 2011). We note that the temperature of the gas in the Sombrero galaxy ($\sim 0.6$ keV $\sim 7 \times 10^6$ K) is consistently recognized as the virial temperature (see also Mathews \& Brighenti 2003). 
On the other hand, recent X-ray observations reveal a variety of the temperature profiles of the hot interstellar medium in early-type galaxies. Diehl \& Statler (2008) categorized the observed temperature profiles into four major groups: isothermal (flat), positive gradient (outwardly rising), negative gradient (falling), and hybrid (falling at small radii and rising at larger radii) (see also Fukazawa et al. 2006). We have, however, a poor understanding the origin and evolution of these temperature gradients. In addition, since X-ray emissivity in the outskirts of the galaxies is very low to be detected with current X-ray satellites, the maximum radius of the observed X-ray emission is usually smaller by a factor of 10 than the virial radius of the CDM halo (Fukazawa et al. 2006). Thus, there is a significant ambiguity keeping the temperature gradient in the outskirts of the galaxies. Under these complicated situations, we focused on the flat temperature profile in this paper.

\subsubsection{Is mass injection along the wind necessary?}
One may think that mass injection along the wind flow due to the SNe is essential and cannot be neglected for plausible galactic wind models. Actually, it may play intrinsic roles as material/energy source of the wind for central star forming region of galaxies that is in a locus of smaller radius than the transonic point. In addition, mass injection results in effective braking force on the wind like gravity. These make essential influences on the wind nature in the central region. 

However, in the resultant wind solution of our analysis for actual CDM halo models, the transonic point forms at far distant region from the galactic center ($\sim 10$ times the scale radius of the CDM halo density distribution that is much larger than the radius of typical locus of the star forming region, see Fig. \ref{fig2}). In such the distant region, we can neglect any effect from star formations, thus can neglect mass injection by SNe along the wind flow when we discuss the acceleration process of the wind. Moreover, when we apply our result to the wind from groups/clusters of galaxies, the mass injection along the flow also can be neglected. Any effect from mass injection is meaningful only at the region in the vicinity of the starting point of the flow, and thus, it does not play any important role on the actual acceleration process of the wind if the transonic point forms sufficiently outside the scale radius as like our case. 

We can conclude that the mass injection along the wind is not necessarily important for galactic wind theory except the case that the transonic point is very close to the galactic center. This simple but analytically effective approximation we adopted in this paper may be allowed as the first step to explore the transonic nature of the galactic winds. 

\subsubsection{Range of the coefficient $K$}
We here discuss the range of expected value of the coefficient $K$ defined in equation \ref{eq:K}. As discussed in Sec. 3, the value of $K$ is critical to determine the topology of the solution curves in the $M^2-x$ phase diagrams (see Figs. 1-5). However, note that the value of $K$ is still uncertain both observationally and numerically. 

Substituting $T=T_{\rm vir}$ into the equation (\ref{eq:KusingT}), we obtain
\begin{equation}
K = f(\alpha) / \gamma.
\end{equation}
Eke, Navarro \& Frenk (1998) adopted $\gamma=2.3$ and Kitayama \& Suto (1997) adopted $\gamma=1.8$ as their canonical value in the analysis of galaxy cluster number counts. For instance, assuming $\gamma=1$, NFW model ($\alpha=1$) yields $K=5.2$ and Fukushige-Makino-Moore model ($\alpha=1.5$) yields $K=2.9$. According to Table \ref{table:f}, a reasonable range is $ 0 < K < 15$ in this case. Each dashed-curve corresponds $K=f(\alpha)/\gamma$ with $\gamma=1, 0.5$ and 0.1 from bottom to top in Fig. {\ref{fig6}.

In addition, the combination of the extra heating mechanisms such as SNe and UV background radiation and the radiative cooling process could keep the temperature to be constant in many situations (see Dekel \& Silk 1986; Yoshii \& Arimoto 1987; Mori, Yoshii \& Nomoto 1999; Babul \& Rees 1992; Efstathiou 1992). In this case, $K$ is no longer the function of the one-parameter family of $\alpha$ quoted above.

\subsection{Implication of the critical condition $\alpha \geq2$}

We have revealed that transonic solution starting from the galactic center needs rather steep density distribution (see Sec. 3.3). 
This is a natural consequence that the solution must approach asymptotically to the Parker's solution in the limit of steep density gradient.
It is striking that none of cosmological $N$-body simulations based on the collisionless CDM predicts such a steep power-law index ($\alpha\geq 2$) around the center of CDM halos. Therefore, the density enhancement of the baryon at the central region of the CDM halo may be essential for the transonic galactic wind from the galactic center. In other words, the gravitational potential induced by a stellar system and/or a central massive black hole plays a crucial role for the acceleration of the galactic wind from the galactic center.

Though, we studied only the effect from the CDM halo as a source of gravitational potential in this paper, it is very interesting to examine the case including the gravitational potential induced by the stellar system and the central massive black hole. We would then also need to consider the effects of the star formation and subsequent feedback process such as stellar winds and SNe heating from massive stars. These feedback processes will supply thermal energy into the gas and will accelerate the outflows.

In a series of forthcoming studies, we plan to report the results taking into account the multiphase states of the gas with cool components in a galaxy, including the effect of the radiative cooling of the gas, energy input from stars and AGN, and the UV background radiation. In this case, the efficiency of the acceleration of the outflow may be quite different.

\section*{Acknowledgments}
This work was supported by the Grant-in-Aid for Scientific Research (A)(21244013), and (C)(20540242, 25400222).

\onecolumn

\appendix

\section{Gauss's hypergeometric function}

Hypergeometric function defines as follows.
\begin{align}
&\,_p\mathrm{F}_q \left[ a_1,\cdots ,a_p;b_1,\cdots,b_q;z \right] =\sum^{\infty}_{n=0} \frac{(a_1)_n \cdots (a_p)_n}{(b_1)_n \cdots (b_q)_n}\frac{z^n}{n!} , \\ \label{eq:hypergeo}
&(a)_n=a(a+1) \cdots (a+n-1)=\frac{\Gamma (a+n)}{\Gamma (a)}, \\
&(a)_0=1.
\end{align}
Especially, in case of $p=2, q=1$, it is called Gauss's hypergeometric function.
It is given by
\begin{equation}
\,_2\mathrm{F}_1 \left[ a,b;c;z \right] =\frac{\Gamma (c)}{\Gamma (a) \Gamma (b)}\sum^{\infty}_{n=0} \frac{\Gamma (a+n) \Gamma (b+n) }{\Gamma (c+n)}\frac{z^n}{n!}. \label{eq:gausshypergeo}
\end{equation}
When using $\alpha$ directly, without substituting a number for it, total mass of galaxy within $x$, $M(x)$ is given only by series expansion, using the Gauss's hypergeometric function. It is given by
\begin{equation}
M(x)= \frac{4\pi \rho_0{r_0}^3}{3-\alpha} x^{3-\alpha} \,_2\mathrm{F}_1 \left[ 3-\alpha,3-\alpha;4-\alpha;-x \right].  \label{eq:jyuryokuG}
\end{equation}
Substituting equation (\ref{eq:jyuryokuG}) in numerator in the right side of equation (\ref{eq:vkou}), we obtain
\begin{align}
(2c_{\rm s}^2/x)-(\textrm{d}\phi /\textrm{d}x) &=\frac{2c_{\rm s}^2}{x}-\frac{4\pi \rho_0{r_0}^2G}{3-\alpha} x^{1-\alpha} \,_2\mathrm{F}_1 \left[ 3-\alpha,3-\alpha;4-\alpha;-x \right], \\
&=\frac{2c_{\rm s}^2}{x}-\frac{4\pi \rho_0{r_0}^2G}{3-\alpha}\frac{\Gamma(4-\alpha)}{\Gamma(3-\alpha)\Gamma(3-\alpha)}\sum^{\infty}_{n=0}\frac{\Gamma(3-\alpha -n)\Gamma(3-\alpha-n)}{\Gamma(4-\alpha-n)}\frac{1}{n!}(-1)^n x ^{n+1-\alpha}, \\
&=\frac{2c_{\rm s}^2}{x} \left[ 1-\sum^{\infty}_{n=0}A_n(-1)^nx^{n+2-\alpha} \right], \\
A_n &=\frac{2\pi \rho_0{r_0}^2G}{c_{\rm s}^2(3-\alpha)}\frac{\Gamma(4-\alpha)}{\Gamma(3-\alpha)\Gamma(3-\alpha)}\frac{\Gamma(3-\alpha -n)\Gamma(3-\alpha-n)}{\Gamma(4-\alpha-n)}.
\end{align}

\label{lastpage}

\end{document}